\documentclass[twocolumn,prl,showpacs,superscriptaddress]{revtex4}  

\usepackage{graphicx}
\usepackage{dcolumn}
\usepackage{bm}
 
\begin{document}  
\def\k {{\bf k}}  
\def\r {{\bf r}}  
\def\R {{\bf R}}  
\def\H {{\bf H}}  
  
\preprint{Taraskin et al.}  
  
\title{Spatial decay of the single-particle density matrix in tight-binding metals: \\  
 analytic results in two  dimensions}

\author{S.~N.~Taraskin}  
 \email{snt1000@cus.cam.ac.uk}  
\affiliation{Department of Chemistry, University of Cambridge,  
             Lensfield Road, Cambridge CB2 1EW, UK}  
 
\author{P.~A.~ Fry}  
 \email{paf25@cam.ac.uk}  
\affiliation{St.John's College, University of Cambridge,  
             St. John's St., Cambridge CB2 1TP, UK}  
  
\author{Xiaodong Zhang}%
 \email{xzhang@helios.phy.ohiou.edu}  
\affiliation{Department of Physics and Astronomy, Ohio  
University, Athens, Ohio 45701 }  
 
\author{D.~A.~Drabold}%
 \email{drabold@ohio.edu}  
 \altaffiliation[Work implemented while on leave at~]{Trinity College, Cambridge}  
\affiliation{Department of Physics and Astronomy, Ohio  
University, Athens, Ohio 45701 }  
  
\author{S.~R.~Elliott}  
\email{sre1@cus.cam.ac.uk} \affiliation{Department of Chemistry,  
University of Cambridge,  
             Lensfield Road, Cambridge CB2 1EW, UK}  
  
\date{\today}
  
\begin{abstract}  
Analytical results for the asymptotic spatial decay of the  
density matrix $\rho({\bf r},{\bf r^\prime})$ in the tight-binding model of the two-dimensional  
metal are presented.   
In various dimensions $D$, it is found analytically and numerically  
that the density matrix decays with distance according to  
the power law, $\rho({\bf r},{\bf r^\prime}) \propto  
|{\bf r}-{\bf r^\prime}|^{-(D+1)/2}$.  
\end{abstract}  
  
 
  
\maketitle  
  
 
The computation of ground-state properties of a  
condensed-matter system  
from the electronic structure depends critically upon the computation of  
the single-particle density matrix (DM).  
If one possesses the occupied  
eigenstates $\psi_i({\bf r})$ of a single-particle Hamiltonian $\hat{\bf H}$, then the  
density matrix at zero temperature can be expressed as  
$\rho({\bf r},{\bf r^\prime}) =\sum_{i~\text{occupied}}  
\psi_i^*({\bf r^\prime}) \psi_i({\bf r})$.  
The diagonal element of the DM, $\rho({\bf r},{\bf r})$, is  
the charge density.   
The electronic energy may be expressed  
as $\text{Tr}(\hat{\bm\rho }\hat{\bf H}$),  
and in a simple case the electronic part of the interatomic forces may be  
expressed as $\text{Tr}(-\hat{\bm\rho }  
\partial\hat{\bf H}/\partial  {\bf R})$, where   
${\bf R}$ is an atomic coordinate.  
Viewed in the position representation, it is clear that  
the decay of $\rho({\bf r},{\bf r^\prime})$ determines  
how {\it locally} one can formulate a calculation of the energy or forces.  
This is of special interest in so-called "quantum order-$N$" methods in modern first-principles computational   
condensed-matter physics \cite{Kohn_96,Goedecker_99:review}. 
 
The DM provides a means to differentiate between a metallic and an insulating state.  
A considerable body of work has been devoted to computing the DM in various systems.   
For insulators, it is well established that, for  
$ |{\bf r}-{\bf r^\prime}| \to \infty $, $\rho ({\bf r},{\bf r^\prime})  
\sim \exp(-\gamma  |{\bf r}- {\bf r^\prime}|)$   
\cite{IsmailBeigi_99,He_01}.  
Recently, we have published  detailed asymptotic  
expansions  for insulators in one, two and three dimensions  
\cite{Taraskin_02:PRL}.   
In {\it metals} the situation is less clear.  
Analytic results are available for  
the free-electron gas in any dimensionality ($D = 1 - 3$)  
 and the DM exhibits a power-law decay  
with Gibbs ringing (from the abrupt cutoff at the Fermi surface at $T=0$)  
\cite{Goedecker_98}.  
Little is known about the ``tight-binding" case,  
except in one dimension where the mathematics is trivial.  One more ``realistic"  
numerical calculation with a density functional Hamiltonian has appeared for Al, which 
produces a DM quite similar to the free-electron gas\cite{Zhang_01}. 
In this Brief Report, we provide analytical asymptotic results for the decay of the DM  
in one dimension and, for the first time, in two dimensions for  
special directions on a square lattice.  
The DM is found to decay generally  as  
$|{\bf r}-{\bf  r^\prime}|^{-(D+1)/2}$ for large $|{\bf r}-{\bf  r^\prime}|$,   
which coincides with the free-electron case.  
Numerics support this law for all three dimensions along   
various lattice directions.  
We provide detailed expressions for  
the decay depending upon the parameters of the one-band 
tight-binding model and the position of the Fermi level.  
The analytic results are  
confirmed by direct numerical evaluation of the DM.  
 
Let us consider a tight-binding Hamiltonian defined on  
a lattice,  
\begin{equation} 
\hat{\bf  H} = \sum_i^N \varepsilon_i |i\rangle\langle i |  
+ \sum_{i\ne j}^N t_{ij} |i\rangle\langle j |~,  
\label{e1} 
\end{equation} 
where the orthonormal site basis $|i\rangle$  
(one electron orbital per site) spans the Hilbert space  
of the state vectors.  
In the case analyzed below, all the site energies,  
$\varepsilon_i = \varepsilon  $, and transfer integrals between nearest neighbours,   
$t_{ij} = t$, are constant  through the lattice.  
The Bloch functions,  
$|{\bf k}\rangle = N^{-1/2} \sum_j \exp\{i{\bf k} \cdot {\bf R_j}\}   
|j\rangle $,  with  dispersion  
$\varepsilon_{\bf k} =  
2t \sum _\alpha^D \cos(k_\alpha a)$ for the simple cubic lattice  
(say, with lattice constant $a=1$ and $\varepsilon = 0$)  
solve the eigenproblem for  
the Hamiltonian (\ref{e1}).  
 
The object to evaluate is the density-matrix   
operator, $\hat{\bm \rho}$, which can be written  
in the momentum  representation as  
$\hat{\bm \rho} = \sum_{\bf k} 
|{\bf k}\rangle f _{\bf k} \langle {\bf k}| $, with   
$ f_{\bf k}$ being the occupation probabilities for  
different eigenstates.  
For an electronic system in thermal equilibrium,  these  
probabilities are the Fermi-Dirac factors,  
$ f_{\bf k} = f(\varepsilon_{\bf k}) = (1+\exp{[(\varepsilon_{\bf k} - \mu)/T]})^{-1} $,  
where $\mu$ is the Fermi level and $T$ is the temperature.  
The matrix elements of the density-matrix operator  
in the site basis,  
$\rho_{ij}= \langle i |\hat{\bm\rho}|j\rangle$,   
 are of special interest for obtaining the decay  
properties of the DM in real space.  
These matrix elements can be written in terms of the  
matrix elements of the Green's function operator,  
$G_{ij}(\varepsilon) = \langle i | 
(\varepsilon - \hat{\bf H})^{-1}|j\rangle $, as:  
\begin{eqnarray} 
\rho_{ij} &=& \frac{1}{N}\int_{-\infty}^{\infty} 
f(\varepsilon)\sum_{\bf k} 
\delta(\varepsilon - \varepsilon_{\bf k})  
\exp\{ i{\bf k} \cdot ({\bf R}_j - {\bf R}_i) \} \text{d}\varepsilon  
\nonumber  
\\ 
&=&  
-\frac{1}{\pi} \text{Im} \int_{-\infty}^{\infty} 
f(\varepsilon) G_{ij}(\varepsilon+i0) \text{d}\varepsilon 
~.   
\label{e2} 
\end{eqnarray} 
The problem of the DM decay is thus partly reduced to the  
problem of the evaluation of the off-diagonal elements  
of the Green's function operator in the site basis, which is  
known to be not at all an easy task  (see e.g. Ref.~\cite{Delves_01} and  
references therein).  
After some standard algebraic manipulations and introducing  
auxiliary integration  (see e.g. \cite{Ehrenreich_76}),    
the expression for $G_{ij} \equiv G_{\nu_\alpha}$ for the simple  
cubic lattice can be  
recast  in the following form:  
\begin{equation} 
G_{\nu_\alpha}(\varepsilon) =  
-\frac{i}{2t} \int_0^\infty \exp\{iz\varepsilon/2t\}  
\prod_\alpha^D i^{\nu_\alpha}J_{\nu_\alpha}(z) 
\text{d}z~,   
\label{e3} 
\end{equation} 
where the integers $\nu_\alpha$ stand for the Cartesian projections of  
the connection vector ${\bf R}_j - {\bf R}_i$, and $J_\nu(z)$ is the Bessel function.   
 
Starting from this point, we are able to proceed 
analytically further  only in the particular case of a square lattice  
($D=2$) along the main diagonal, $\nu_x = \nu_y\equiv \nu$  
(and in the one-dimensional case as well),  
when the integral over $z$ can be taken exactly  
\cite{Gradshteyn_00:book}:  
\begin{eqnarray}  
& & \int_0^\infty  \exp\{iz\varepsilon/2t\}  
J_\nu^2(z) \text{d}z=  
\nonumber 
\\  
& & \frac{1}{\pi}Q_{\nu -1/2}(1-2\epsilon^2) +  
\frac{i}{2}P_{\nu -1/2}(1-2\epsilon^2)~,  
\label{e4} 
\end{eqnarray} 
if the energy belongs to the band region,  
$0<|\epsilon|<1$, where $\epsilon \equiv \varepsilon/4t$.  
The functions $P_\nu$ and $Q_\nu$  are the associated  
Legendre functions of  the first and second kind, respectively.  
The expression for the DM can then be recast as  
\begin{equation}  
\rho_\nu  =  
(-1)^{\nu}\frac{2}{\pi} \int_{-1}^{\epsilon_{\text{F}}}  
Q_{\nu -1/2}(1-2\epsilon^2) \text{d}\epsilon~,  
\label{e5} 
\end{equation} 
where $\epsilon_{\text{F}} \equiv \mu/4t$ is the  
dimensionless Fermi level and the zero-temperature  
case is implied.   
If the Fermi level lies above the band, i.e.  
$\epsilon_{\text{F}} \ge 1 $, all the states are occupied  
at zero temperature and the DM is $\rho_\nu = \delta_{\nu 0}$,  
just reflecting the completeness of the basis set.  
This property, together with the even character of the integrand  
in Eq.~(\ref{e5}), allows us to rewrite the expression  
for the DM (for $\nu > 0$) in the following form:  
\begin{equation}  
\rho_\nu  =  
(-1)^{\nu}\frac{2}{\pi} \int_{0}^{\epsilon_{\text{F}}}  
Q_{\nu -1/2}(1-2\epsilon^2) \text{d}\epsilon~,  
\label{e6} 
\end{equation} 
and consider for definiteness only $\epsilon_{\text{F}} > 0$.  
 
The integral in Eq.~(\ref{e6}) can be  simplified in the  
asymptotic limit of large $\nu \to \infty$ by using  
the asymptotic expression for  
$Q_{\nu-1/2}(\cos\phi)$ \cite{Gradshteyn_00:book},  
\begin{equation}  
Q_{\nu-1/2}(\cos\phi) \simeq  
\sqrt{\frac{\pi}{2\nu\sin\phi}}  
\cos\left( \nu\phi +\frac{\pi}{4} \right)~,  
\label{e7} 
\end{equation} 
where $\phi = \cos^{-1}(1-2\epsilon^2)$, so that  
\begin{equation}  
\rho_\nu  \simeq  
\frac{(-1)^{\nu}}{2\pi^{3/2}\nu^{1/2}}  \int_{0}^{\phi_0}  
\sqrt{ \frac {\sin\phi}{1-\cos\phi}}  
\cos\left( \nu\phi +\frac{\pi}{4} \right) \text{d}\phi~,  
\label{e8} 
\end{equation} 
with $\phi_0 = \cos^{-1}(1-2\epsilon^2_{\text{F}})$.  
The final step revealing the explicit asymptotic  
dependence of the DM on $\nu$ can be made if the  
Fermi level lies not far from the midband point  and $\phi_0 \ll 1$.  
In that case, the first term of the integrand in Eq.~(\ref{e8})  
can be expanded in $\phi$, resulting in  
\begin{equation}  
\rho_\nu  \simeq  
(-1)^{\nu}\frac{\sqrt{2}}{\pi^{3/2}\nu} \int_{0}^{\psi_0}  
\cos\left(\psi^2 +\frac{\pi}{4} \right)\text{d}\psi~,  
\label{e9} 
\end{equation} 
with $\psi_0=\sqrt{\nu\phi_0}$.  
All the non-trivial dependence on $\nu$ is now in the  
upper limit of the integral, the latter being  proportional  
to  the Fresnel integrals $C(x)$ and $S(x)$,  
\begin{equation}  
\rho_\nu  \simeq  
(-1)^{\nu}\frac{1}{\pi\nu\sqrt{2}} 
\left[  
C(\sqrt{ 2\nu\phi_0/\pi}) - S(\sqrt{ 2\nu\phi_0/\pi}) 
\right]  
~,  
\label{e10} 
\end{equation} 
and in the asymptotic limit, $\sqrt{ 2\nu\phi_0/\pi} \to \infty$   
\cite{Gradshteyn_00:book},  
\begin{equation}  
\rho_\nu  \simeq  
(-1)^{\nu}\frac{1}{(\pi\nu)^{3/2}\sqrt{2\phi_0}} 
\cos\left(\nu\phi_0 - \frac{\pi}{4}\right)~.   
\label{e11} 
\end{equation} 
As follows from Eq.~(\ref{e11}), the DM decays according to the power law,  
$\rho_{ij}\propto |{\bf R}_j -{\bf R}_i|^{-3/2}$, at least along  
the main diagonal in the square lattice.  
 
\begin{figure} 
\centerline{\includegraphics[width=7cm,angle=270]{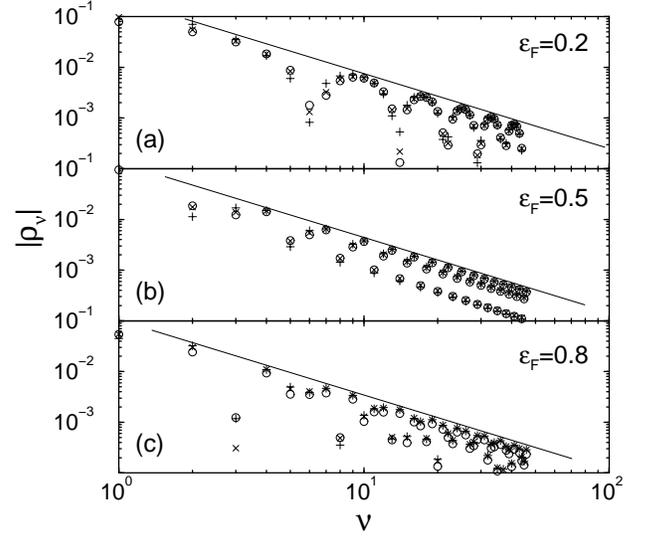}}  
\caption{  
The dependence of the absolute value of the DM, $|\rho_{\nu}|$,  
on the lattice index $\nu $ along the main diagonal  
$[1,1]$ in the square lattice (${\bf R}_j -{\bf R}_j =  
(\nu_x, \nu_y) \equiv (\nu, \nu)$) for a  
tight-binding model of a crystalline metal at zero temperature  
and various positions of the Fermi-level,  
$\epsilon_{\text{F}}=\mu/4t$, as marked in (a)-(c).  
The open circles represent the exact numerical result obtained  
from Eq.~(\ref{e2}).  
The crosses and pluses correspond to the  
approximate results obtained by using  
Eqs.~(\ref{e9}) and (\ref{e11}), respectively.  
The straight solid lines  show the power-law  
dependence, $\rho_\nu \propto \nu^{-3/2}$.  
}  
\label{f1} 
\end{figure}  
 
\begin{figure} 
\centerline{\includegraphics[width=7cm,angle=270]{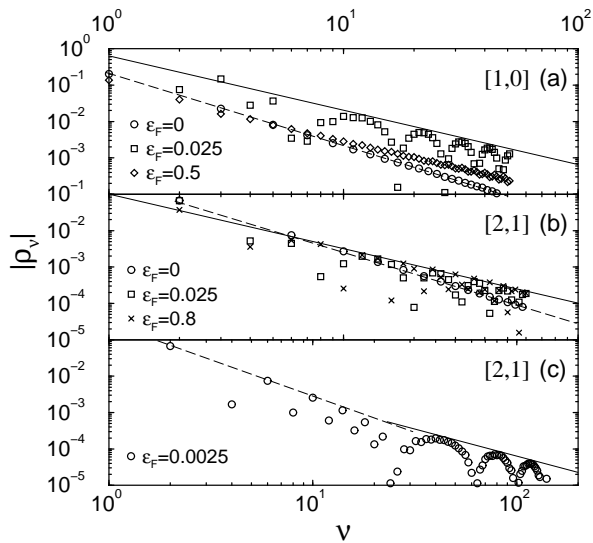}}  
\caption{  
The dependence of the absolute value of the DM, $|\rho_{\nu}|$,  
on the lattice index $\nu $: (a) along the direction   
$[1,0]$ in the square lattice (${\bf R}_j -{\bf R}_j =  
(\nu_x, 0) \equiv (\nu, 0)$)  
and (b)-(c)  
along the direction   
$[2,1]$ in the square lattice (${\bf R}_j -{\bf R}_j =  
(2\nu_y, \nu_y) \equiv (2\nu, \nu)$) 
for the same model as in Fig.~\ref{f1}.  
The results are obtained numerically  
from Eq.~(\ref{e2}).  
The straight solid and dashed lines  show the power-law  
dependences, $\rho_\nu \propto \nu^{-3/2}$  and  
$\rho_\nu \propto \nu^{-2}$, respectively.  
}  
\label{f2} 
\end{figure}  
 
All the analytical results presented above can be verified  
by direct numerical analysis.  
In Figs.~\ref{f1}(a)-(c), we show the dependence of  
the DM versus $\nu$ along the diagonal $[1,1]$ in the square  
lattice for different positions of the Fermi level.  
The exact numerical results (open circles) have been  
obtained by both direct summation  over the first  
Brillouin zone in Eq.~(\ref{e2}) (over $10^8$ points) and  
by integration of Eq.~(\ref{e5}) (both methods give identical  
results).  
The approximate analytic results according to  
Eqs.~(\ref{e9}) and (\ref{e11}) are given by the crosses   
and pluses, respectively.  
Good agreement between the exact and approximate  
dependencies is evident, even for the relatively  
large values of the Fermi-level position  far away from the midband  
region (see Fig.~\ref{f1}(c)).  
The solid straight lines in Fig.~\ref{f1}  
 corresponding to the power law,  
$|\rho_\nu | \propto \nu^{-3/2}$,  
confirm the same law for the DM decay.  
 
In order to verify the power-law decay of the DM along other  
directions in the simple square lattice, we have calculated  
the DM numerically for these directions and have 
found the same asymptotic behavior,  
$\rho_{\nu} \propto \nu^{-3/2}$.  
The results for directions $[1,0]$ and $[2,1]$ 
and different positions of the Fermi level  
are presented in Figs.~\ref{f2}(a)-(c).   
 
For the special case that the Fermi level lies exactly 
at the band center, $\epsilon_{\text{F}}=0$, note that 
for the main diagonal $[1,1]$, the DM vanishes for $\nu >0$  
(see Eq.~(\ref{e6})). 
This behavior is also manifested in other directions:  
for $\epsilon_{\text{F}} \approx 0$ from  
$\rho_\nu \propto \nu^{-3/2} $ to  
$\rho_\nu \propto \nu^{-2} $ (see the dashed lines in  
Figs.~\ref{f2}(a)-(b)).  
Fig.~\ref{f2}(c) shows how this occurs for the direction  
$[2,1]$ when $\epsilon_{\text{F}} \rightarrow 0$.

The analysis of the DM decay  in different dimensions  
can be performed analytically for $D=1$ and numerically  
for $D=3$.  
First, we look at the one-dimensional  system.  
In the case of zero temperature, the integrals in  
Eq.~(\ref{e2}) can be taken exactly, resulting  
in the following expression for the DM ($|\mu/2t|\le 1$):  
\begin{equation}  
\rho_\nu  =   
\frac{1}{\pi\nu} 
\sin\left[\nu\left( \sin^{-1}\frac{\mu}{2t} +\frac{\pi}{2} \right)\right]~.   
\label{e12} 
\end{equation} 
This expression has the correct limits for a fully occupied  
($\mu =2t$) and an empty ($\mu =-2t$) band, viz.  
$\rho_\nu = \delta_{\nu0}$ and $\rho_\nu = 0$, respectively,  
and shows the power-law decay,  
$\rho_\nu \propto \nu^{-1}$.  
Bearing in mind  the decay law, $\rho_\nu \propto \nu^{-3/2}$,  
found for $D=2$, we can infer that the generalized  
law in all dimensions is:   
\begin{equation}  
\rho_{ij}=\rho_\nu   \propto \nu^{-(D+1)/2}  
\propto |{\bf R}_j - {\bf R}_i|^{-(D+1)/2}~.   
\label{e13} 
\end{equation} 
%
%
 
\begin{figure} 
\centerline{\includegraphics[width=7cm,angle=270]{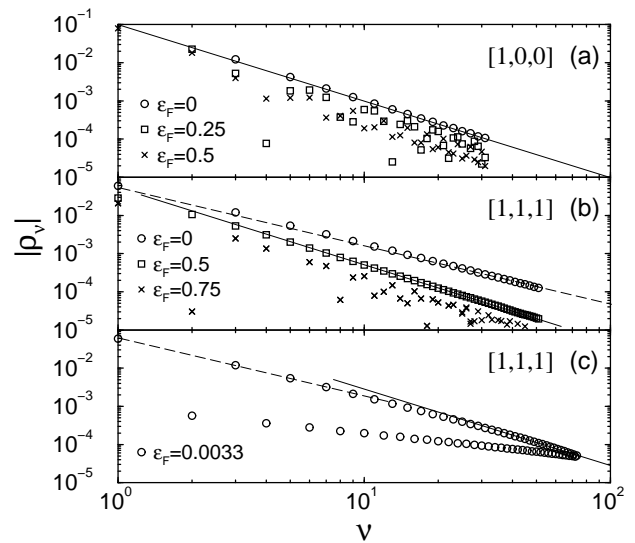}}  
\caption{  
The dependence of the absolute value of the DM, $|\rho_{\nu}|$,  
on the lattice index $\nu $: (a) along  the direction   
$[1,0,0]$ in the simple cubic lattice (${\bf R}_j -{\bf R}_j =  
(\nu_x, 0, 0) \equiv (\nu, 0, 0)$)  
and (b)-(c)  
along  the direction   
$[1,1,1]$ in the simple cubic lattice (${\bf R}_j -{\bf R}_j =  
(\nu_x, \nu_y, \nu_z ) \equiv (\nu, \nu,\nu)$)  
for a  
tight-binding model of a crystalline metal at zero temperature  
and various positions of the Fermi level,  
$\epsilon_{\text{F}}=\mu/8t$, as marked.  
The results are obtained numerically  
from Eq.~(\ref{e2}).  
The straight solid and dashed lines  show the power-law  
dependences, $\rho_\nu \propto \nu^{-2}$ and $\rho_\nu \propto \nu^{-3/2}$, respectively.  
}  
\label{f3} 
\end{figure}  
 
In order to check Eq.~(\ref{e13}) for $D=3$,  
we have calculated numerically the DM  for the simple  
cubic lattice along different directions for various positions of the  
Fermi level.  
The results are shown in Figs.~\ref{f3}(a)-(c),  
from which it is clear that indeed the DM asymptotically satisfies  
Eq.~(\ref{e13}) and  
$\rho_{\nu} \propto \nu^{-2}$.    
 
Similar to the $2D$-case, the mid-band location  
of the Fermi energy brings additional symmetry to the problem  
which can change the asymptotic behaviour of the DM  
along certain directions.  
For example, if $\epsilon_{\text{F}}=0$ then $\rho_\nu =0$ 
for $\nu >0$ along the direction  
$[1,1,0]$ and $\rho_\nu \propto  \nu^{-3/2}$ along the $[1,1,1]$  
direction (see the dashed line in Fig.~\ref{f3}(b);  
Fig.~\ref{f3}(c) demonstrates how the new asymptotic behavior appears  
when the Fermi level approaches the midband position).

\begin{figure} 
\centerline{\includegraphics[width=8.5cm]{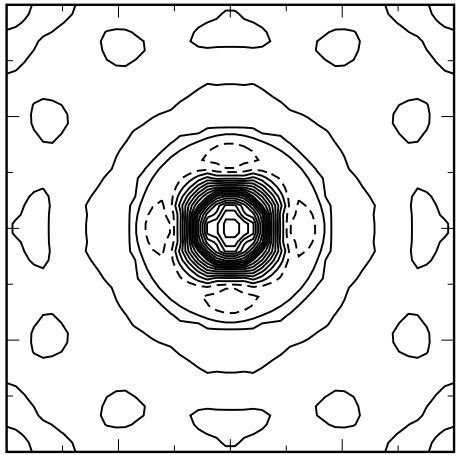}}  
\caption{  
Contour plot of the real-space density matrix for Al calculated   
 in the $\{100\}$ plane 
for the conventional cubic unit cell 
(the $x-y$ axes are parallel to the bonds).
}  
\label{f4} 
\end{figure}  

\begin{figure} 
\centerline{\includegraphics[width=6.2cm,angle=270]{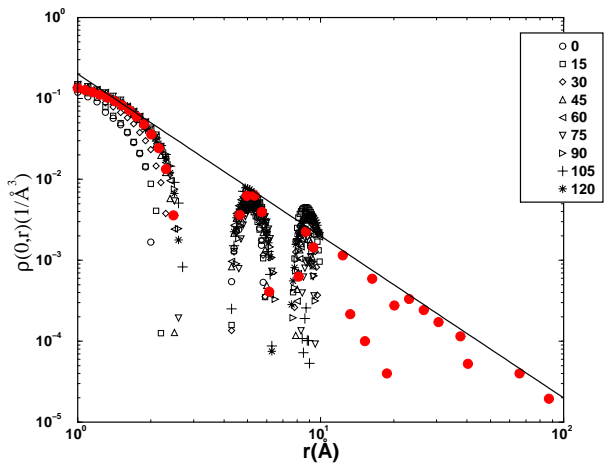}}  
\caption{  
Spatial decay  of the real-space density matrix for Al calculated for  
different angular directions, as marked 
(zero angle refers to the bond direction), in the $\{100\}$ plane 
for the conventional cubic unit cell. The solid circles represent 
calculated values of the density matrix for the free-electron gas model 
with the same electron density as for Al. The solid line shows 
an $r^{-2}$ dependence. 
}  
\label{f5} 
\end{figure}  

The above analytical and numerical results have been obtained 
for very simple tight-binding models. 
This of course leaves open the question about the generality of 
our findings. 
In order to answer this question, at least to some extent, 
we have calculated the single-electron DM for a realistic model 
of fcc aluminum ($500$ atom supercell with the box side 
of $20.25 \AA$) using an approximate density functional Hamiltonian 
in the local density approximation (see Ref.~\cite{Zhang_01} for 
more detail). 
The results are presented in Figs.~\ref{f4} and \ref{f5}. 
The real-space contour plot for the DM in the $\{100\}$ 
plane for the conventional cubic unit cell is shown in Fig.~\ref{f4}. 
From this plot, we can see  the isotropic 
metallic nature of the bonding, in contrast to the case of 
semiconductors with covalent bonding 
(cf. Fig.~3 in Ref.~\cite{Zhang_01}).  
The spatial decay of the DM along different directions 
in the same symmetry plane for Al is shown in Fig.~\ref{f5},  
together with the data  calculated for the free-electron gas 
model \cite{Goedecker_98} (with the electron density being the same 
as that for Al, i.e. $0.185\AA^{-3}$). 
It is clearly seen that the DM for Al decays in a very similar 
fashion to that for the free-electron gas model, i.e. 
$\rho \propto r^{-2}$. 
Therefore, these results support the generality of our model 
calculations.

In  conclusion, we have presented analytical and numerical  
arguments supporting the power-law decay of the  
density matrix,  
$ \rho_{ij}  
\propto |{\bf R}_j - {\bf R}_i|^{-(D+1)/2}$, in  
tight-binding models of metals in different dimensions  
at zero temperature.  
The main result is  the analytical asymptotic dependence  
of the density matrix versus distance along the main diagonal  
in the square lattice (see Eq.~(\ref{e11})).   
Apparently, the sharp cutoff  induced by the Fermi-Dirac  
distribution at zero temperature in the integration over the  
energy spectrum, independently of the shape of 
the density of states, results in the power-law decay of the density  
matrix in crystalline metals.

DAD thanks the National Science Foundation for support 
under grants DMR 0081006 and DMR 0205858. 
 

\end{document}